# Sources of Variability in Alpha Emissivity Measurements at LA and ULA Levels, a Multicenter Study.

Brendan D. McNally[a*], Stuart Coleman[a], William K. Warburton[a], Jean-Luc Autran[b], Brett M. Clark[c], Jodi Cooley[d], Michael S. Gordon[e], and Zhengmao Zhu[f]

[a]*XIA LLC, 31057 Genstar Rd., Hayward, CA 94544, USA*

[b]*Aix-Marseille Marseille University & CNRS, Faculté des Sciences – Service 142, F-13397 Marseille Cedex 20, France*

[c]*Honeywell, 15128 E. Euclid Ave., Spokane, WA 99216 USA USA*

[d]*Southern Methodist University – Physics Department, Dallas, TX 75275 USA*

[e]*IBM TJ Watson Research Center, 1101 Kitchawan Rd., Yorktown Heights, NY 10598 USA*

[f]*IBM SRDC, 2070 Rte 52, Bldg 300, 1W4-111, Hopewell Junction, NY 12533 USA*

**Abstract**

Alpha emissivity measurements are important in the semiconductor industry for assessing the suitability of materials for use in production processes. A recently published round-robin study that circulated the same samples to several alpha counting centers showed wide center-to-center variations in measured alpha emissivity. A separate analysis of these results hypothesized that much of the variation might arise from differences in sample-to-entrance window separations. XIA recently introduced an ultra low background counter, the UltraLo-1800 ("UltraLo"), that operates in a fundamentally different manner from the proportional counters used at most of the centers in the original study. In particular, by placing the sample within the counting volume, it eliminates the sample-to-entrance window separation issue noted above, and so offers an opportunity to test this hypothesis. In this work we briefly review how the UltraLo operates and describe a new round-robin study conducted entirely on UltraLo instruments using a set of standard samples that included two samples used in the original study. This study shows that, for LA ("Low Alpha" between 2 and 50 $\alpha$/khr-cm$^2$) sample measurements, the only remaining site-to-site variations were due to counting statistics. Variations in ULA ("Ultra-Low Alpha" < 2 $\alpha$/khr-cm$^2$) sample measurements were reduced three-fold, compared to the earlier study, with the measurements suggesting that residual activity variations now primarily arise from site-to-site differences in the cosmogenic background.

*Keywords:* Alpha counting, alpha particles, ionization detector, low background, soft errors

## 1. Introduction

Almost all materials emit alpha particles, primarily from the decay of trace uranium and/or thorium impurities, but also from exposure to radon or other radioactive contaminants. The ability to measure alpha particle emissions from materials is therefore important in applications that are especially sensitive to alpha particles, including dark matter detection and other rare-event physics experiments, environmental monitoring, and the semiconductor industry. In the latter, alpha particles can cause soft errors [1] and controlling them is critical to improving the reliability of electronic devices operated on or near the earth's surface. Emissivity measurements are therefore used by manufacturers in the semiconductor industry both to determine the suitability of materials for use in devices and to monitor process contamination levels. When coupled with accelerated alpha tests [2], emissivity measurements are also used to estimate a portion of the devices' soft error rate (SER), and, more generally, anticipated product reliability.

In 2010, the Alpha Consortium formed to assess the accuracy and reliability of alpha emissivity measurements, supposing that, at sample activity levels many times higher than background (e.g. LA, or "Low Alpha", between 2 and 50 $\alpha$/khr-cm$^2$), measurement agreement between sites should be feasible, while at the lowest activity levels (e.g. ULA, or "Ultra-Low Alpha", < 2 $\alpha$/khr-cm$^2$) instrument backgrounds might introduce site dependent variability. A round-robin study was organized to test this proposal, with a set of two LA and two ULA samples sent randomly to a group of nine participating counting centers. The study was blinded and anonymized so that the participants did not know which samples they were measuring or which samples the other participants measured. The study results [3], however, unexpectedly showed that emissivity values measured at the LA level varied by factors of two or more that correlated with the site, while the ULA samples' measurement uncertainties were so large that it was not possible to determine if their values showed similar site-to-site variations. The authors hypothesized that the observed variability at the LA level could be explained by differences between the counters' low energy discriminator settings, while variability at the ULA level was probably affected both by these settings and by the counters' different backgrounds. A follow up study using both an LA sample and a calibrated point source [4], was made to explicitly determine the role of low energy discriminator settings on site-to-site measurement variability. Those results indicate that it plays a minor role, if any, and continued to show the factor of two site-to-site measurement variability, even with the calibrated, NIST-traceable alpha point source. The authors then suggested a different source of site-to-site variation – differences in the separation between samples and counter entrance windows, which cause variations in counter efficiency because alpha particles emitted at low angles are less likely to enter the counter as the distance to the entrance window increases.

---

[*] Corresponding author. Tel.: +1-510-401-5760; fax: +1-510-401-5761; e-mail: brendan@xia.com.

Both the Alpha Consortium and the industry in general are concerned by the inability of the counting centers to produce site-to-site reproducible measurements at LA levels because it calls into question their underlying accuracy and reliability. XIA, a participant in the Alpha Consortium experiments, has developed a new alpha-particle counter (UltraLo-1800, hereafter "UltraLo") whose design eliminates the entrance window, and has demonstrated the ability to measure samples at the sub-ULA ($< 1$ $\alpha$/khr-cm$^2$) level [5]. While XIA was the only participant to use the UltraLo in both Alpha Consortium studies, enough are now in use to allow the Alpha Consortium's round-robin experiment to be repeated with the new instruments to test this hypothesis.

In this work, we describe and analyze the results of a round-robin field assessment conducted between six alpha counting sites using the UltraLo instrument. We will explore how much variation is observed between participants in the measurement of a set of standard samples and determine the degree to which observed variations are random or systematic by participant.

## 2. Methods

The UltraLo is a windowless instrument, where samples are inserted directly into a gas-filled, large-area active counting chamber. In this system, samples are arranged on a tray that is then moved into measurement position via an electromechanical stage, thereby ensuring repeatable positioning from measurement to measurement. The UltraLo operates in a fundamentally different manner from the gas proportional counters predominately used in the Alpha Consortium experiments. In essence, it is an ionization chamber without internal gain whose geometry intentionally exaggerates differences between signals from alpha particles generating ionization tracks that originate from its different surfaces [6]. When an ionization track is detected in the chamber, the resultant signal waveforms are captured by onboard digital electronics and their pulse shapes are analyzed to determine the location from which the track originated. Figure 1 shows a schematic cross section of the UltraLo's ionization chamber, and signal waveforms typical of tracks originating from the sample under test and from two interior chamber surfaces. As may be seen, the waveforms differ significantly in both amplitude and risetime between these three cases. The UltraLo's software recognizes these differences and actively rejects non-sample tracks, thereby driving background rates an order of magnitude or more lower than can be achieved in the best commercially available gas proportional counters [7]. To reduce sensitivity to cosmogenic events, additional software tests detect and reject tracks that span the entire chamber or appear to originate above the surface of the sample tray.

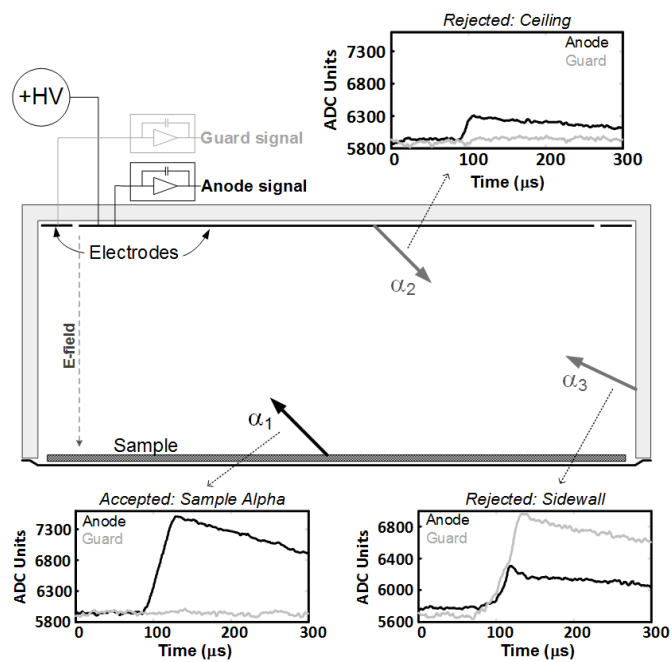

Figure 1: Cross sectional schematic view of UltraLo-1800 chamber, with representative signal waveforms from 3 alpha events. $\alpha_1$ is an alpha particle emitted from the sample. $\alpha_2$ represents a background alpha from the ceiling of the chamber, and $\alpha_3$ denotes a background alpha from the sidewall of the chamber

Three samples were used in this study, two of which were also used in the first round of the Alpha Consortium measurements [3]. These samples were chosen because calibration standards at the desired activity levels do not exist and they had both been thoroughly measured as part of the original Alpha Consortium study. Further, for over two years these samples have also been used during pre-shipment counter qualification measurements at XIA and their activity levels have been found to be stable and reproducible. The first was the aluminum alloy sample labeled "LA-1", and the second was the titanium sample labeled "ULA-2" [3]. The third sample, which is new to this study, was a bare 300 mm diameter silicon wafer whose activity level is expected to be nominally zero. It was supplied by author M.S.G at IBM and is labeled "SULA-Si-1" (SULA standing for "sub-ULA").

This study reports on the results from 6 participants at 6 separate locations, including XIA. XIA participated twice, making the first and last measurements of each sample. After an initial measurement at XIA, the sample set was shipped sequentially to each participant, with instructions to prepare and measure each sample following a detailed procedure (see supporting online materials). The procedure defined the sample handling and preparation process, as well as counter configuration and measurement parameters. In particular, before measuring the ULA samples, participants were instructed to clean the samples in a prescribed manner to remove any contamination caused by prior handling or shipping and to arrange the sample pieces on the counting tray in a consistent manner. Following measurement by the 5th participant the samples were remeasured at XIA in order to assess whether or not their activity levels might have changed over the course of the study.

Participants were also instructed to configure their counting systems in an identical manner, which entailed selecting the same electrode configuration, setting the same trigger thresholds – which serve as a proxy for the low energy discriminator – and using prescribed measurement times. The measurement times were: 24 hours for sample LA-1, and 168 hours for samples ULA-2 and SULA-Si-1. Following the measurements, participants submitted their raw waveform data to XIA for analysis (see supporting online materials). To preclude operator prejudice from possibly biasing results, a standardized analysis was applied to all data sets that consisted of the following steps. First, the initial 48 hours of data were removed from the 168 hour measurements of samples ULA-2 and SULA-Si-1, a procedure that has been shown to be necessary to allow incidental radon contamination to decay away, and any residual moisture from the washing procedure to evaporate [5]. The LA-1 data were analyzed as received as its activity rate was known to be several times higher than the anticipated worst-case radon contamination rates. Second, the raw waveform data from each data set were processed using the standard analysis package CounterMeasure supplied with the UltraLo. For the analyses, CounterMeasure was configured with identical settings except for two instrument calibration parameters: gain correction and risetime cut. These were set to the values supplied by the manufacturer for each specific instrument, which are also the values that the instrument normally uses as installed at its counting center. For each analysis completed, CounterMeasure reported a measured activity level and standard deviation, which are the values reported here (see supporting online materials).

In the CounterMeasure analyses the detected alpha-particle counts are assumed to be from the stochastic process of radioactive decay, and thus drawn from a Poisson distribution characterized by the count rate $\mu = N / t$, where N is the number of alpha counts detected, and t is the time interval of observation (e.g., measurement time). In the results that follow, the reported emissivities are then the observed count rates $\mu$ scaled by the counter efficiency and divided by the area of the sample in the counter's active region; the uncertainties reported are from counting statistics alone (i.e., $\sigma = \sqrt{N}/t$), similarly scaled and divided.

## 3. Results and observations

### 3.1. Sample Stability

In the data presented in the following sections, location 1 and location 7 represent the 'before' and 'after' measurements conducted at XIA. For all three samples, these results are within 1 standard deviation of each other, supporting the assertion that the samples' activities were stable throughout the duration of the study.

### 3.2. Sample 1 – Aluminum Alloy LA-1

The measurement results from sample LA-1 are shown in Figure 2 and summarized in Table I. Good agreement among the participants is evident; 6 of the 7 values are within $1\sigma$ of their mean value of 38.0 $\alpha$/khr-cm$^2$, while location 3 records a value just outside of this limit, approximately $1.09\sigma$ from the mean. The complete set of values is fully consistent with a normal distribution of measurement data. The relative uncertainty (defined here as $\sigma/\mu$, where $\mu$ is the observed emissivity) for all measurements is approximately 4%. There is no sign of the site-to-site variations seen in the previous round-robin studies.

Table I: Summary of emissivity results for Sample LA-1

| Location | Emissivity ($\alpha$/khr-cm$^2$) | Error ($\alpha$/khr-cm$^2$) |
|---|---|---|
| 1 | 37.67 | 1.54 |
| 2 | 38.65 | 1.57 |
| 3 | 36.35 | 1.51 |
| 4 | 38.83 | 1.57 |
| 5 | 38.25 | 1.56 |
| 6 | 36.86 | 1.53 |
| 7 | 39.36 | 1.58 |

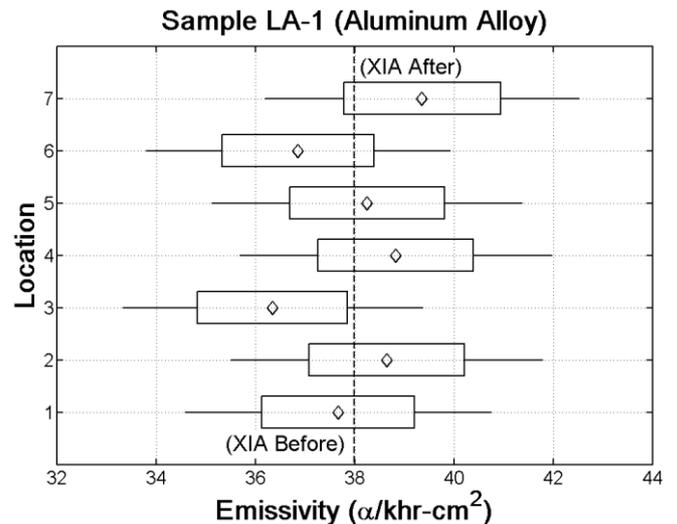

Figure 2: Emissivity measurements for Sample LA-1. The diamonds represent the measured emissivities, the boxes bound +/- 1 sigma, the whiskers bound +/- 2 sigma. The mean value of 38.00 $\alpha$/khr-cm$^2$ is depicted by the vertical dashed line.

Table II: Summary of emissivity results for Sample ULA-2

| Location | Emissivity ($\alpha$/khr-cm$^2$) | Error ($\alpha$/khr-cm$^2$) |
|---|---|---|
| 1 | 0.74 | 0.10 |
| 2 | 1.52 | 0.14 |
| 3 | 0.91 | 0.11 |
| 4 | 0.80 | 0.10 |
| 5 | 0.73 | 0.10 |
| 6 | 0.81 | 0.11 |
| 7 | 0.92 | 0.11 |

Table III: Summary of emissivity results for Sample SULA-Si-1

| Location | Emissivity ($\alpha$/khr-cm$^2$) | Error ($\alpha$/khr-cm$^2$) |
|---|---|---|
| 1 | 0.70 | 0.10 |
| 2 | 1.19 | 0.13 |
| 3 | 0.47 | 0.08 |
| 4 | 0.44 | 0.08 |
| 5 | 0.35 | 0.07 |
| 6 | 0.42 | 0.08 |
| 7 | 0.69 | 0.10 |

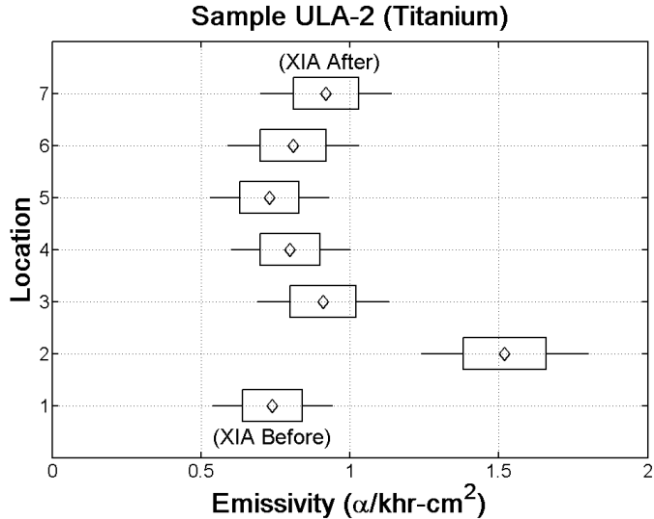

Figure 3: Emissivity results for Sample ULA-2. The diamonds represent the measured emissivities, the boxes bound ± 1 sigma, the whiskers bound ± 2 sigma.

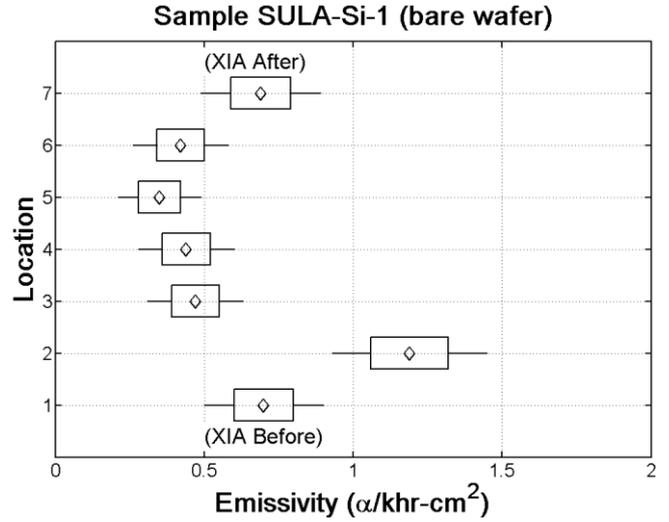

Figure 4: Emissivity results for Sample SULA-Si-1. The diamonds represent the measured emissivities, the boxes bound ± 1 sigma, the whiskers bound ± 2 sigma.

### 3.3. Sample 2 – Titanium ULA-2

Emissivity measurement results from sample ULA-2 are shown in Figure 3 and summarized in Table II. Six of the seven measurement results are within 1$\sigma$ of each other, while the value measured at location 2 (1.52) is approximately 5$\sigma$ away from the mean value of the other 6 locations (0.82). All 7 measurements are spread over a range of 0.79 $\alpha$/khr-cm$^2$, and the average relative uncertainty for these 120 hour measurements is 12%.

### 3.4. Sample 3 – Bare Wafer SULA-Si-1

Figure 4 presents the measurement results from sample SULA-Si-1, and it is immediately evident that location 2 has again measured a value significantly different from the other locations. The results show 4 out of 7 locations measuring a value within 1$\sigma$ of each other (mean value = 0.42). Locations 1 and 7 measure a value approximately 2.8$\sigma$ above these values (mean value = 0.70), while location 2 measures a value 5.9$\sigma$ above these values (1.19). In total, the data report a spread of 0.84 $\alpha$/khr-cm$^2$, and an average relative uncertainty of 14%. All results for sample SULA-Si-1 are summarized in Table III.

## 4. Discussion

### 4.1. Low Alpha Measurements (Sample LA-1)

The variability in the LA-1 results observed in this study stands in marked contrast to the variability observed from the same sample in the first Alpha Consortium experiment. For comparison, the LA-1 results from both the Alpha Consortium experiment and this study are overlaid in Figure 5.

The uncertainties reported for both the Alpha Consortium and the present study are based solely on counting statistics, and in the former, they are clearly unable to account for the observed variability in the measurements. This indicates that additional sources of error must be present but unrecognized, the most likely being differences in separation between samples and entrance windows. In most cases, the magnitude of the reported uncertainty scales with the square root of sample counting time as one might expect, noting that the times displayed in Figure 5 are total counting time which most, but not all, participants split evenly between sample and background measurements. The 6 and 20.5 hour measurements in the first study at location L1.J2 and L1.J1 were XIA's values, also taken with the UltraLo, and are in statistical agreement with the values obtained in the present work 26 months later. Within the precision of the current set of measurements (± 1.5 $\alpha$/khr-cm$^2$), the observed variations in emissivities obtained with the UltraLo appear to be fully described by counting statistics and also to be stable with time.

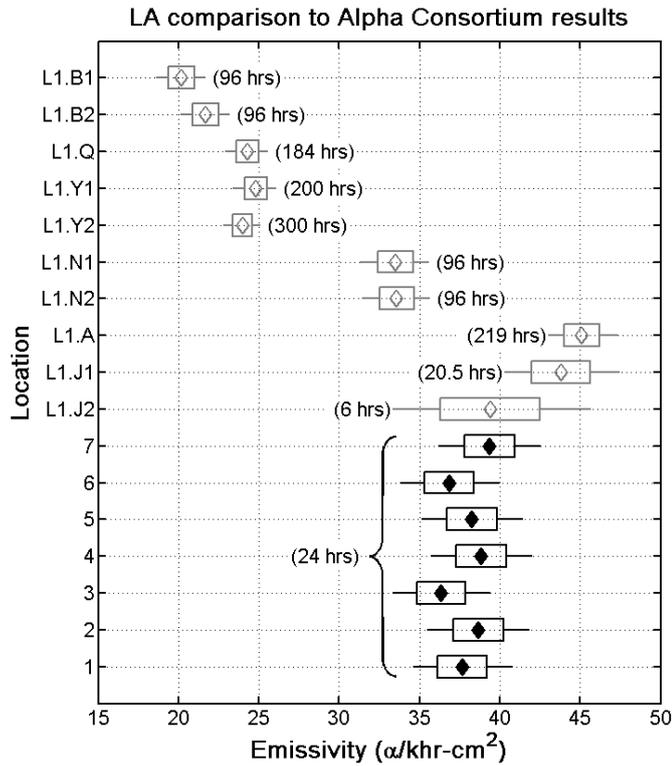

Figure 6: Comparison of sample LA-1 results from this study (black) to Alpha Consortium results of same sample (grey) [3]. Total counting time (background + gross sample) is noted in parentheses for Alpha Consortium results.

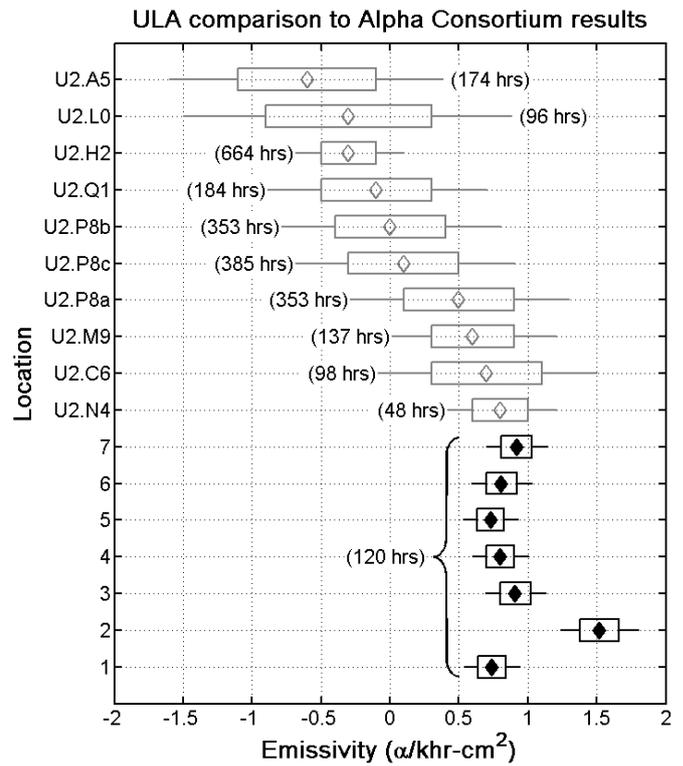

Figure 5: Comparison of sample ULA-2 results from this study (black) to Alpha Consortium results of same sample (grey) [3]. Total counting time (background + gross sample) is noted in parentheses for each Alpha Consortium result.

Further, the lack of site-to-site variations in the UltraLo LA-1 sample results is consistent with the hypothesis that the variations observed in the Alpha Consortium study arose from inconsistent sample-to-window positioning, which the new instrument's design eliminates. This in turn suggests that it may then be possible to also achieve site-to-site uniformity using windowed gas proportional counters, provided that a method for ensuring a consistent sample-to-window separation can be developed and uniformly employed at all counting centers.

*4.2. Ultra Low Alpha Measurements (Samples ULA-2 & SULA-Si-1)*

Figure 6 compares the measurement results of sample ULA-2 from both the Alpha Consortium study and this one. It is immediately clear that only positive values appear in the current study, as opposed to some negative values reported for ULA-2 in the Alpha Consortium measurements. Negative emissivity values are unphysical, and can either result from counting statistics (e.g. a value of 0.5 measured with a standard deviation of 2.0 will often produce values less than 0) or result from some error in the background correction procedure. A typical such error, particularly for gas proportional counters of the type predominately used in the Alpha Consortium study derives from the failure of the assumption that the insertion of a sample does not perturb the background counting rate. If this assumption is not met, then subtracting the measured 'background" rate from the measured "sample + background" rate does not produce the "sample" rate because the "background" rates in the two measurements are different. At the ULA level it is quite difficult to assure that this criterion is actually met. For example, a small amount of residual activity on the surface of the empty sample tray may be subsequently covered by a lower activity sample, so that the measured "background" is higher than during the "sample + background" measurement, which results in a negative emissivity value being calculated. The UltraLo does not employ a background subtraction step and therefore is not sensitive to this class of error. However, omitting this step might lead to a small systematically positive bias if any background counts are misidentified by the UltraLo's electronic discrimination system. A discussion of how large this bias might be can be found at the conclusion of the discussion section.

In the first study, the 48 hour measurement at location U2.N4 was XIA's value, again taken with the UltraLo, and is in good agreement with the values measured at XIA – locations 1 and 7 – in the present study 27 months later. While some site-to-site variability is still present, its variance is over three times lower in this study. In addition, by reducing the uncertainties reported in the Alpha Consortium experiment by a factor of 3.5, we are now able to explore possible mechanisms for the remaining site-to-site variability.

Figure 7 compares the measurement results from samples ULA-2 and SULA-Si-1. Recalling that ULA-2 is a titanium sample that is expected to have some residual alpha activity while SULA-Si-1 is an ultra-pure silicon sample with nominally zero activity, several observations may be made. First, even excluding locations 1, 2 and 7 (i.e. XIA and location 2), it is clear that a non-zero activity level is consistently reported for SULA-Si-1 (average value 0.42 ± 0.8), which implies that either the silicon activity is greater than zero or else a background contribution exists that the UltraLo's pulse shape analysis algorithms do not identify. Secondly, ULA-2's values are consistently higher than those from SULA-Si-1 (average difference 0.31), suggesting that the titanium sample is indeed slightly more active than the silicon sample. The data indicate that the UltraLo can reliably detect this difference in a 168 hour measurement even though both values are well below the value of 2 $\alpha$/khr-cm$^2$ that defines the ULA limit. In particular, this difference is still well resolved at location 2, even with its increased overall counting levels.

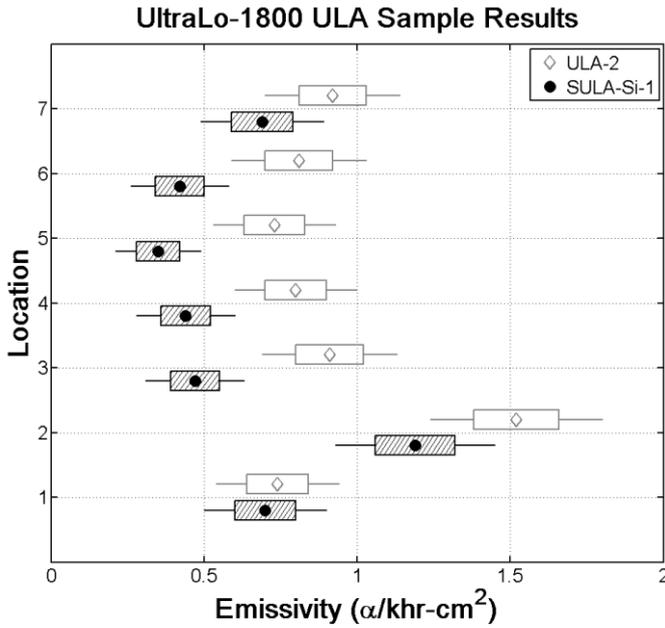

Figure 7: Comparison of emissivity results from ULA samples used in this study. The grey diamonds represent the measured emissivity of sample ULA-2, and the black circles represent the measured emissivity of sample SULA-Si-1. In both cases, the boxes bound ± 1 sigma, the whiskers bound ± 2 sigma.

Our third observation is that location 2 consistently measures higher activity levels than the other locations. While this might be an instrument issue rather than a location issue, ULA-2 was also measured in this instrument for 48 hours at location 1 as part of a set of final calibration measurements prior to delivering it to location 2. The calibration value of 0.93 ± 0.17 $\alpha$/khr-cm$^2$, relative to 1.52 ± 0.14 $\alpha$/khr-cm$^2$ reported in this study, suggests that the location 2 environment is the more likely source of the observed increased counting rates. The most likely candidate for this increase is the cosmogenic background, which is known to produce measurable alpha counts in gas-filled counters, including the UltraLo [8], with rates that vary with both altitude and measurement site shielding.

To estimate this effect, the study participants were asked to provide a description of the facility where their instrument was located, including elevation and facility construction. From this information we constructed a simple 'overburden' model to estimate the mass thickness above the instrument at each location. The results of this survey and calculation are summarized in Table IV, where it is immediately evident that location 2 has significantly less overburden than the other sites.

Table IV: Location description and estimated overburden. Note that overburden estimates include both atmospheric and facility contributions.

| Loc | Description | Est. Overburden (g cm$^{-2}$) |
|---|---|---|
| 1,7 | 1st floor lab, single story building, no shielding | 2069 |
| 2 | 1st floor lab, single story building, no shielding | 1991 |
| 3 | 1st floor lab, 3 story building, concrete cons. | 2169 |
| 4 | Basement lab, underground, concrete cons. | 2195 |
| 5 | Basement lab, 3 story building, concrete cons. | 2233 |
| 6 | Basement lab, 3 story building, concrete cons. | 2225 |

Figure 8 plots emissivity measurement results from sample SULA-Si-1 versus estimated site overburden, where we see that emissivities indeed drop as overburden increases. The displayed fit uses equation 4 from [9], which describes the cosmogenic neutron flux dependence on atmospheric depth and has the form

$$F_{ALT}(d) = \exp\left[\frac{(1033.7 - d)}{131.3}\right], \qquad (1)$$

where the fixed terms 1033.7 g cm$^{-2}$ and 131.3 g cm$^{-2}$ are taken from [9], and represent the sea-level atmospheric overburden and neutron attenuation length, respectively. Considering Figure 8, we see that this model also explains why the XIA measurements (locations 1 & 7) are also consistently high for sample SULA-Si-1. However, while this exponential fit to the estimated overburdens is evocative, the overburden estimates are not sufficiently accurate at this point to ascribe specific meanings to the fitting parameters. The observed correlation does strongly suggest that a residual cosmogenic background term is the sole remaining source of site-to-site measurement differences made at the ULA level using the XIA counter. Based on the physics of their operation, proportional counters will also be sensitive to cosmogenic events. However, such issues as whether event rates are constant in time, depend upon the presence or absence of a sample, etc., and hence can be adequately corrected by the background measurement are too complex to be dealt with within the scope of this paper. Further, the magnitude of uncertainties reported in the Alpha Consortium measurements preclude drawing any conclusions based on the reported values themselves. This same cosmogenic background term may also explain some, or perhaps all, of the residual activity found for sample SULA-Si-1, whose activity was expected to be zero. However, further research, perhaps including underground measurements, will be needed to distinguish between the cosmogenic background and other possible residual terms such as radon emitted from the materials used to construct the counters.

Considering Figure 8, and the relationship between overburden and the SULA-Si-1 measurement data, several observations can be made about the potential for systematic bias in our measurement data. First, it seems clear that, for all sites, some emissivity contribution arises from cosmogenics and results in a site-specific positive bias to any UltraLo measurement. The relative contributions of cosmogenics, radon, or other residual terms to the observed bias cannot be determined without conducting a long underground measurement. However, our data do provide two upper bound estimates for these contributions. First, the SULA-Si-1 measurement at each site provides an upper bound on the total possible positive bias at that site. Second, the lowest SULA-Si-1 measurement in Figure 8 (site 5) provides an upper bound on the radon or other residual source contribution to the bias.

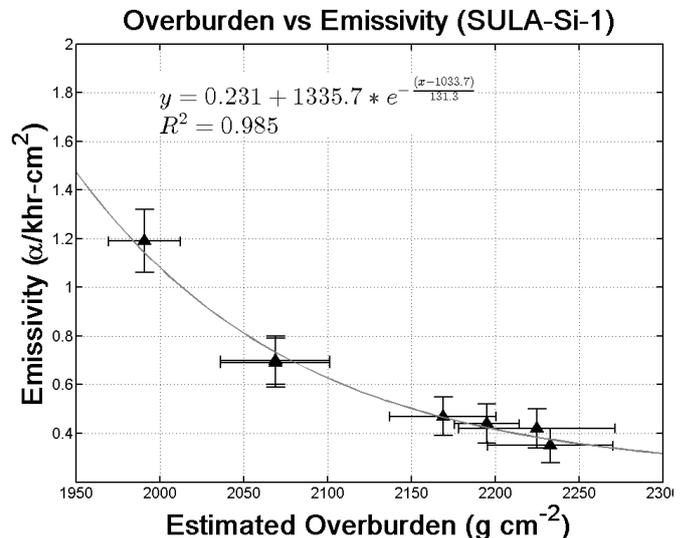

Figure 8: Estimated overburden vs Emissivity for Sample SULA-Si-1

In commercial practice, material suppliers generally certify products at two activity grades: LA ($< 50$ $\alpha$/khr-cm$^2$) and ULA ($< 2$ $\alpha$/khr-cm$^2$). When certifying LA material, this bias – typically less than 1 $\alpha$/khr-cm$^2$, as inferred from the SULA-Si-1 data – represents a negligible source of measurement error. However, when certifying ULA material, this effect cannot be ignored, and demonstrates the importance of correctly characterizing this contribution when estimating the true activity of ULA samples.

## 5. Conclusions

The results of this study are encouraging. In contrast to the Alpha Consortium measurements, the data presented here show minimal site-to-site variability for LA level emissivity measurements using the new instrumentation. The variability that is observed is within bounds explained by counting statistics. This suggests that it may be possible to also achieve site-to-site uniformity using windowed gas proportional counters, provided that the community can agree on a standard sample-to-window separation and develop a practical methodology to repeatedly achieve it, particularly in the presence of sample-to-sample height variations.

At the ULA level, the large uncertainties reported for nearly all results from the Alpha Consortium study make it difficult to deduce specific sources of the observed variation. When making emissivity measurements of a sample that is at or below an instrument's background rate, an accurate determination of counter background is critical in order to produce a statistically supportable result. In the absence of a true "zero-emissivity" sample that can be placed at a standard sample-to-window separation, it appears difficult to design a standardized background measurement for windowed gas proportional counters that is conceptually capable of producing a result at the accuracy needed for ULA measurements that are reproducible site-to-site. While measurement data at ULA levels using the new UltraLo instrument still exhibit site-to-site variations, these have been reduced to about 0.8 $\alpha$/khr-cm$^2$, a factor of about 3 smaller than in the earlier work. We hypothesize that these remaining variations are primarily due to a class of background events in the counter, of cosmogenic origin, that cannot yet be fully identified and rejected by the XIA pulse shape analysis software. XIA is currently working with the analysis software to try to quantify and mitigate this effect. Further research will be required to determine if these cosmogenic terms are stable in time, in which case they might be subtracted as a background term. Underground measurements will be necessary to determine whether other background components (such as internally generated radon which could vary by instrument) are also present at the ULA level.

## Author Contributions

## Acknowledgments

The authors wish to acknowledge the generous counting time provided by each of the participating institutions. Further, the authors wish to acknowledge numerous fruitful discussions with the members of the Alpha Consortium; particularly Jeff Wilkinson, who served as a valuable sounding board while discussing and reviewing early results from this work.